\documentclass[aps,pra,graphics,twocolumn,floatfix,notitlepage,superscriptaddress]{revtex4-2}
\usepackage[english]{babel}
\usepackage{textcomp,xcolor,natbib}
\usepackage{dcolumn}
\usepackage{graphicx}
\graphicspath{{Figures/}}

\usepackage{amsmath, bbm}
\usepackage{latexsym}
\usepackage{amsfonts}   
\usepackage{amssymb}
\usepackage{array}      
\usepackage{epsfig}
\usepackage{txfonts}
\usepackage{xcolor,braket}
\usepackage[colorlinks=true,linkcolor=blue,urlcolor=blue,citecolor=blue]{hyperref}
\usepackage[normalem]{ulem}
\usepackage{soul}
\usepackage{dsfont}

\usepackage{xfrac}  
\usepackage{relsize}

\usepackage{mathrsfs}

\usepackage{cancel}
\usepackage{enumitem}

\usepackage{xspace}

\newcommand{\ketbra}[1]{\ensuremath{\ket{#1}\bra{#1}}}

\newcommand{\id}{\ensuremath{\mathbbm{1}}}\newcommand{\1}{\id}
\newcommand{\abs}[1]{\lvert #1 \rvert}
\newcommand{\norm}[1]{\lVert #1 \rVert}
\newcommand{\tr}{\ensuremath{\operatorname{tr}}}
\newcommand{\pd}{\ensuremath{\mathscr H_{\mathcal J_t}}}
\renewcommand{\Re}{\operatorname{Re}}

\newcommand{\asq}{\abs\alpha^2}
\newcommand{\RO}{\ensuremath{\Psi\text{-}R}}

\begin{document}

\title{Generalized Rate Operator Quantum Jumps via Realization-Dependent Transformations}

\author{Federico Settimo}
\email{federico.f.settimo@utu.fi}
\affiliation{Department of Physics and Astronomy,
University of Turku, FI-20014 Turun yliopisto, Finland}

\author{Kimmo Luoma}
\affiliation{Department of Physics and Astronomy,
University of Turku, FI-20014 Turun yliopisto, Finland}

\author{Dariusz Chru\'sci\'nski}
\affiliation{Institute of Physics, Faculty of Physics, Astronomy and Informatics,
Nicolaus Copernicus University, Grudziadzka 5/7, 87-100 Toru\'{n},
Poland}

\author{Bassano Vacchini}
\affiliation{Dipartimento di Fisica ``Aldo Pontremoli'', Universit{\`a} degli Studi di Milano, Via Celoria 16, I-20133 Milan, Italy}
\affiliation{Istituto Nazionale di Fisica Nucleare, Sezione di Milano, Via Celoria 16, I-20133 Milan, Italy}

\author{Andrea Smirne}
\affiliation{Dipartimento di Fisica ``Aldo Pontremoli'', Universit{\`a} degli Studi di Milano, Via Celoria 16, I-20133 Milan, Italy}
\affiliation{Istituto Nazionale di Fisica Nucleare, Sezione di Milano, Via Celoria 16, I-20133 Milan, Italy}

\author{Jyrki Piilo}
\affiliation{Department of Physics and Astronomy,
University of Turku, FI-20014 Turun yliopisto, Finland}

\begin{abstract}
    The dynamics of open quantum systems is often solved by stochastic {unravelings} where the average over the state vector realizations reproduces the density matrix evolution. We focus on quantum jump descriptions based on the rate operator formalism. In addition to displaying and exploiting different equivalent ways of writing the master equation, we introduce state-dependent rate operator transformations within the framework of stochastic pure state realizations, allowing us to extend and generalize the previously developed formalism. As a consequence, this improves the controllability of the stochastic realizations and subsequently greatly benefits when searching for optimal simulation schemes to solve open system dynamics. At a fundamental level, intriguingly, our results show that it is possible to have positive {unravelings} --  without reverse quantum jumps and avoiding the use of auxiliary degrees freedom --  in a number of example cases even when the corresponding dynamical map breaks the property of P-divisibility, thus being in the strongly non-Markovian regime. 
\end{abstract}

\maketitle

\section{Introduction}
\label{sec:intro}
Stochastic unravelings are a powerful tool to describe the dynamics of open quantum systems \cite{Breuer-oqs, Rivas-Huelga-OQS}.
With this formalism, the time evolution of the state of the system is described as the average over different realizations of a stochastic process on the set of quantum states.
Such stochastic processes can be separated in two major families: they can be either diffusive \cite{Percival-qsd, Barchielli-traj, Diosi-NMQSD, Yu-NMQSD-PRL, Wiseman-diffusive-unr, Luoma-diffusive-NMQJ, Viviescas-entanglement-diffusive, Caiaffa-W-diffusive} or the deterministic evolution can be interrupted by sudden discontinuous jumps \cite{Plenio-jumps-review, Dalibard-MCWF, Daley-jumps-many-body, Dum-MonteCarlo-emission, Dum-MonteCarlo-optics, Gambetta-jump-nM, Guevara2020, Perfetto2022, Chiriaco-many-body-jumps}.
In this paper, we will focus on the latter situation.
These quantum jump methods are particularly convenient for simulating high-dimensional master equations and have been linked to several distinct experimental scenarios \cite{Basche-jump-spectroscopic-obs, Peil-cyclotron-jump, Jelezko-defect-jump-spectroscopy, Gleyzes-jump-recording, Vijay-jumps-obs-superconducting, Minev-jump-mid-flight}.

The standard jump unraveling method for Markovian dynamics, namely the Monte-Carlo wave function (MCWF), consists of jumps whose effects and probabilities are fixed directly by the rates and operators in the master equation \cite{Plenio-jumps-review, Dalibard-MCWF}.
The probabilities of the jumps are guaranteed to be positive if and only if all rates are positive, with the MCWF method failing whenever this condition is not satisfied.
The positivity of all rates is equivalent, under suitable assumptions of regularity, to the completely-positive(CP)-divisibility of the dynamical map \cite{BLPV-colloquium}, meaning that the dynamics can be arbitrarily subdivided in intermediate completely positive maps.
The notion of CP-divisibility has been connected to the definition of Markovianity for open system dynamics \cite{RHP, rivas-quantum-nm}.

The MCWF method has been generalized to non-positive rates by the non-Markovian quantum jumps (NMQJ) method \cite{Piilo-NMQJ-PRL, Piilo-NMQJ-PRA}.
However, the different stochastic realizations are no longer independent, thus making the simulations more expensive.
Nevertheless, it was shown that it is possible to generalize the MCWF method and maintaining independent realization also under the weaker assumption of positive(P)-divisibility of the dynamical maps \cite{Smirne-W}.
This method relies on the definition of the rate operator (RO) \cite{Diosi-orthogonal-jumps, Diosi-stochastic-repr, Diosi-stochastic-state-reduction, Diosi2017} and was therefore named rate operator quantum jumps (ROQJ).
Violations of P-divisibility have also been linked to a different definition of non-Markovianity \cite{BLPV-colloquium, Wißmann-BLP-P-div}.
Interestingly, these unravelings rely on jumps to mutually orthogonal states and have been linked to the study of pointer bases \cite{Busse2010a, Diosi2000}.

Recently, in \cite{Chruscinski-Quantum-RO}, the RO formalism has been expanded by employing the non-unique ways of writing the master equation {by applying arbitrary time-dependent transformations}, thus leading to the formulation of a family of distinct RO unravelings for the same master equation.
In this paper, we generalize those results by allowing {the arbitrary transformation} to depend on the current state of the particular realization.
This generalized RO not only enhances the efficiency of controlling the stochastic realizations, but also enables the characterization of certain dynamics violating P-divisibility, thus non-Markovian according to all definitions, while preserving the independence between the different realizations.
Noticeably, we are able to do so without requiring any additional ancillary degree of freedom.

The rest of the paper is organized as follows.
In Sec.~\ref{sec:RO}, we recall the main features of open system dynamics and of quantum jump unravelings, focusing on the rate operator formalism.
In Sec.~\ref{sec:generalized-RO}, we introduce the generalized RO, by allowing it to depend on the current state of the stochastic realization.
This generalized RO is characterized in Sec.~\ref{sec:characterization}, with a particular focus on the conditions for its positivity for all realizations.
In Sec.~\ref{sec:examples}, we present some examples showing the flexibility given by this approach.
Importantly, we also show that it is possible to have an unraveling with a positive RO for all realizations even for some dynamics which break P-divisibility.
Finally, we present the conclusions of our work in Sec.~\ref{sec:conclusions}.

\section{Rate operator formalism}
\label{sec:RO}
We start by recalling the open quantum systems formalism and the quantum jump unravelings to describe open system dynamics, focusing in particular on the ROQJ formalism.
\subsection{Open quantum systems}
The time-local master equation governing the time evolution of a finite-dimensional open quantum system can be written as $d\rho/dt = \mathcal L_t[\rho]$, with the generator $\mathcal L_t$ given by \cite{Gorini1976, Lindblad1976}
\begin{equation}
    \label{eq:ME}
    \mathcal L_t[\rho] = - i[H(t),\rho] + \sum_\alpha\gamma_\alpha(t)L_\alpha(t)\rho L_\alpha^\dagger(t)-\frac12\big\{\Gamma(t),\rho\big\},
\end{equation}
where $H(t) = H(t)^\dagger$ is the system Hamiltonian, $L_\alpha(t)$ are the Lindblad operators with rates $\gamma_\alpha(t)$, and $\Gamma(t) = \sum_\alpha\gamma_\alpha(t)L_\alpha^\dagger(t)L_\alpha(t)$.
The rates $\gamma_\alpha(t)$ can be temporarily negative, with the {dynamical map} $\Lambda_t=T\exp\left(\int_0^td\tau\,\mathcal L_\tau\right)$ being completely positive \cite{Breuer-oqs, Chruscinski-nM-local, Chruscinski2022}.
Positivity of the rates, however, is equivalent to CP-divisibility of the dynamical map, i.e. $\forall t\ge s\ge0$ it is possible, under suitable regularity conditions \cite{BLP-PRA}, to decompose $\Lambda_t = \Lambda_{t,s}\Lambda_s$ for completely positive operators $\Lambda_{t,s}$.
Simple positivity of $\Lambda_{t,s}$, on the other hand, corresponds to a P-divisible dynamical map, which is equivalent to \cite{BLPV-colloquium, Kossakowski-necessary}
\begin{equation}
    \label{eq:P-div-general}
    \sum_\alpha\gamma_\alpha(t)\abs{\braket{\varphi_i|L_\alpha(t)|\varphi_j}}^2\ge0
\end{equation}
for all orthonormal bases $\{\varphi_i\}_i$, for all $i\ne j$.

A common way to look at the master equation \eqref{eq:ME} in the context of quantum unravelings is to write it as the sum of a jump term
\begin{equation}
    \label{eq:jump_ME}
    \mathcal J_t[\rho] \coloneqq \sum_\alpha\gamma_\alpha(t)L_\alpha(t)\rho L_\alpha^\dagger(t)
\end{equation}
and a driving term
\begin{equation}
    \label{eq:driving_ME}
    \mathcal D_t[\rho] \coloneqq - i(K(t)\rho - \rho K^\dagger(t)),
\end{equation}
with the effective non-Hermitian Hamiltonian
\begin{equation}
    \label{eq:K_effective-Hamiltonian}
    K(t) \coloneqq H(t) - \frac i2 \Gamma(t).
\end{equation}
On the other hand, such a decomposition is highly non-unique.
In fact, any transformation \cite{Chruscinski-Quantum-RO}
\begin{gather}
    \label{eq:transf_J}
    \mathcal J_t[\rho]\mapsto\mathcal J_t^\prime[\rho] \coloneqq \mathcal J_t[\rho] + \frac12(C(t)\rho + \rho C^\dagger(t))\\
    \label{eq:transf_K}
    K(t)\mapsto K^\prime(t) \coloneqq K(t)-\frac i2 C(t),
\end{gather}
for an arbitrary operator $C(t)$, preserves Eq.~\eqref{eq:ME}.
Such a freedom in writing the master equation is different from the conventional approach employed in MCWF methods, which relies on the invariance of the master equation under unitary transformations on the set of Lindblad operators \cite{Breuer-oqs}.

\begin{table*}
    \begin{tabular}{ |c|c|c|c|c|c| } 
     \hline 
     {} & MCWF & $W$-ROQJ & $R$-ROQJ & {$\Psi$}-ROQJ & NMQJ \\
     \hline
     CP-divisible & \checkmark & \checkmark & \checkmark & \checkmark & \checkmark \\
     \hline 
     P- but not CP-divisible & $\times$ & \checkmark & $\circ$ & \checkmark & \checkmark$^*$ \\
     \hline
     Non-P-divisible & $\times$ & $\times$ & $\times$ & $\circ$ & $\circ$ \\
     \hline
     Independent realizations & \checkmark & \checkmark & \checkmark & \checkmark & $\times$ \\
     \hline
    \end{tabular}
    \caption{Comparison of the {$\Psi$}-ROQJ unravelings with the previously introduced methods; \checkmark: yes, $\times$: no, $\circ$: sometimes, $^*$: yes, unless some rates are negative since the beginning of the dynamics.
    Each column refers to a distinct unraveling method, the first three rows denote the applicability of the methods to different classes of dynamics, while the fourth row denotes the possibility to realize each trajectory independently from the others.
    We note in particular that {$\Psi$}-ROQJ is the only method that can tackle also some non-P-divisible-dynamics, while keeping the different realizations independent, as shown in Sec. \ref{subsec:non-P-div}.
    As explained in the text, the {$\Psi$}-ROQJ includes the $W$- and the $R$-ROQJ as special cases, while there is no inclusion among the latter two.
    The table refers to ROQJ methods without reversed jumps; note that all the ROQJ methods can be supplemented with reverse jumps \cite{Piilo-NMQJ-PRL}, leading to correlated realizations and extending their range of applicability to non-P-divisible dynamics.
    }
    \label{tab:RO_comparisons}
\end{table*}

\subsection{Quantum jump unravelings}
In the literature, there have been introduced many different unravelings consisting of piecewise deterministic processes on the set of pure states on the system's Hilbert space $\mathscr H$.
The exact dynamics of Eq.~\eqref{eq:ME} is reproduced by averaging over all stochastic realizations, with the form of the deterministic and jump process that can vary significantly for the different unraveling methods.
Differently from other methods, our approach does not use additional degrees of freedom \cite{Imamoglu-stochastic, Garraway-nonperturbative-decay, Garraway-decay-atom-coupled, Breuer-double-Hilbert, Breuer-trajectories-nM, Becker2023}, which would require additional computational effort, nor temporarily negative probabilities for the occupation of certain states \cite{Muratore-Ginanneschi-influence-martingale}.

Whenever CP divisibility holds, it is possible to unravel the dynamics via the MCWF method \cite{Dalibard-MCWF, Plenio-jumps-review}, with deterministic evolution
\begin{equation}
    \label{eq:MCWF-det}
    \ket{\psi(t)}\mapsto\ket{\psi(t+dt)}=\frac{(\1-iK(t)dt)\ket{\psi(t)}}{\norm{(\1-iK(t)dt)\ket{\psi(t)}}},
\end{equation}
where $K(t)$ is the effective non-Hermitian Hamiltonian of Eq.~\eqref{eq:K_effective-Hamiltonian}, interrupted by sudden jumps
\begin{equation}
    \label{eq:MCWF-jump}
    \ket{\psi(t)}\mapsto\ket{\psi(t+dt)} = \frac{L_\alpha(t)\ket{\psi(t)}}{\norm{L_\alpha(t)\ket{\psi(t)}}}
\end{equation}
with probability
\begin{equation}
    \label{eq:P_jump_MCWF}
    {p^\alpha_{\psi(t)}} = \gamma_\alpha(t)\norm{L_\alpha(t)\ket{\psi(t)}}^2dt.
\end{equation}
Naturally, this method calls for the positivity of all rates $\gamma_\alpha(t)$.

The requirement of positivity of all rates can be weakened by considering the ROQJ formalism.
One possible way to do so consists of unraveling with jumps to the eigenstates of the operator \cite{Smirne-W, Diosi-orthogonal-jumps}
\begin{equation}
    \label{eq:W}
    {W_{\psi(t)} \coloneqq \big(\1-P_{\psi(t)}\big)\mathcal J_t[P_{\psi(t)}]\big(\1-P_{\psi(t)}\big),}
\end{equation}
where $P_\psi=\ketbra\psi$, with probabilities given by the corresponding eigenvalues multiplied by the infinitesimal time increment $dt$.
Following the nomenclature of \cite{Chruscinski-Quantum-RO}, we call this unraveling method W-ROQJ, emphasising that the jumps and their probabilities are fixed by the eigenstates and eigenvalues of $W$.
We further note that the pre-jump state {$\psi(t)$} is an eigenstate of {$W_{\psi(t)}$}, so that the state after the jump is always orthogonal to {$\psi(t)$}.
The deterministic evolution is generated by the non-Hermitian non-linear effective Hamiltonian {$K^W_{\psi(t)} = K(t)+\Delta_{\psi(t)}$}, with
\begin{equation}
    {\Delta_{\psi(t)} = \frac i2\sum_\alpha\gamma_\alpha(t)\left(2L_\alpha(t)\ell^*_{\psi(t),\alpha}(t)-\abs{\ell^*_{\psi(t),\alpha}(t)}^2\right),}
\end{equation}
where {$\ell_{\psi(t),\alpha}(t) = \braket{\psi(t)|L_\alpha(t)|\psi(t)}$}.
The operator {$W_{\psi(t)}$} does not depend on the particular form \eqref{eq:transf_J} of the master equation \cite{Diosi-stochastic-repr, Diosi-stochastic-state-reduction} and is positive for all states $\psi$ if and only if the dynamics is P-divisible \cite{Caiaffa-W-diffusive}.
Therefore, such method can be used to unravel any P-divisible dynamics, significantly beyond the range of applicability of the MCWF.

The ROQJ method can be extended \cite{Chruscinski-Quantum-RO} relying on the non-uniqueness of the decomposition of the master equation \eqref{eq:transf_J}-\eqref{eq:transf_K} and considering jumps $\ket{\psi(t)}\mapsto\ket{{\varphi_{\psi(t)}^j}}$ to the eigenstates of the RO
\begin{equation}
    \label{eq:RO}
        {R_{\psi(t)} \coloneqq \mathcal J_t^\prime[P_{\psi(t)}] = \mathcal J_t[P_{\psi(t)}] + \frac12(C(t)P_{\psi(t)} + P_{\psi(t)} C^\dagger(t))}
\end{equation}
with probability
\begin{equation}
    \label{eq:p_jump_RO}
    {p^j_{\psi(t)} = \lambda^j_{\psi(t)}dt}
\end{equation}
given by the corresponding eigenvalue multiplied by the time increment $dt$.
We refer to this method as the $R$-ROQJ.
The deterministic evolution is as in Eq.~\eqref{eq:MCWF-det}, but using the transformed non-Hermitian Hamiltonian $K^\prime(t)$ of Eq.~\eqref{eq:transf_K}.
The non-uniqueness of the unravelings could allow one to design different realizations for the stochastic process, possibly simplifying the computational task of simulating the dynamics.
Whenever P-divisibility holds, the RO can have at most one negative eigenvalue and the existence of at least one positive RO as in Eq.~\eqref{eq:RO} is guaranteed by the dissipativity of the dynamics, a stronger requirement than P-divisibility \cite{Chruscinski-Quantum-RO}.

Going beyond the MCWF, it is possible to deal with temporarily negative rates by using the NMQJ technique as in \cite{Piilo-NMQJ-PRL,Piilo-NMQJ-PRA}, by considering reverse jumps $\ket{\psi_i(t)}=L_\alpha(t)\ket{\psi_j(t)}/\norm{L_\alpha(t)\ket{\psi_j(t)}}\mapsto\ket{\psi_j(t)}$, with probability depending on the {ratio} $N_j(t)/N_i(t)$ between the occupations of $\ket{\psi_j(t)}$ and $\ket{\psi_i(t)}$.
{However, one needs to know the average state $\rho(t)$ to compute the ratios, and therefore the realizations become dependent one on the other, making the simulation more expensive.}
The same method can be employed similarly also for the RO whenever some eigenvalues are negative.
If, instead, the {eigenvalues} of the RO, and therefore the jump probabilities, are positive for all realizations, we say that such unraveling is a positive unraveling.
{In this case, the stochastic realizations are independent, since the jump probabilities can be calculated from the state of the given individual realization, and the simulation is more efficient.}

\section{Generalized Rate Operator}
\label{sec:generalized-RO}
We now proceed to generalize the ROQJ method by exploiting the invariance under Eqs.~\eqref{eq:transf_J} and \eqref{eq:transf_K} on each individual trajectory.
This will significantly extend the class of generated unravelings of a given master equation with respect to both $W$-ROQJ and $R$-ROQJ, which will indeed be regained as special instances.

Suppose that, at some time $t$, the state can be written in terms of the single realizations as $\rho=\sum_ip_i(t){P_{\psi_i(t)}}$.
From the point of view of a single realization {$\ket{\psi(t)}$}, it is possible to choose the transformation $C$ of Eq.~\eqref{eq:transf_J}-\eqref{eq:transf_K} {to depend not only on time but also} {\it on the current state} {$\ket{\psi(t)}$} and leaving the average evolution unaffected.
The new generalized RO is therefore of the form
\begin{equation}
    \label{eq:state-dep-RO}
    {\Psi\text{-}R_{\psi(t)}\coloneqq\mathcal J_t[P_{\psi(t)}] + \frac12\big(C_{\psi(t)}P_{\psi(t)} + P_{\psi(t)} C_{\psi(t)}^\dagger(t)\big),}
\end{equation}
with {$C_{\psi(t)}$} that can depend non-trivially on the current state of the realization {$\psi(t)$}.
To emphasize the dependence on the state {$\psi(t)$}, we refer to the generalized RO as {$\Psi$}-ROQJ.
Indeed, whenever {$C_{\psi(t)}$} does not depend on {$\psi(t)$}, the {$\Psi$}-ROQJ reduces to the $R$-ROQJ of Eq.~\eqref{eq:RO}.
The deterministic evolution is non-linear because of the state-dependence of the effective non-Hermitian Hamiltonian
\begin{equation}
    \label{eq:state-dep-K}
    {K_{\psi(t)} \coloneqq H(t)-\frac i2\Gamma(t) - \frac i2C_{\psi(t)}.}
\end{equation}

Since {$C_{\psi(t)}$} appears only to {applied to $\ket{\psi(t)}$}, it is possible to simplify the transformation by defining the unnormalized vector
\begin{equation}
    \label{eq:Phi}
    {\ket{\Phi_{\psi(t)}}\coloneqq C_{\psi(t)}\ket{\psi(t)}.}
\end{equation}
This way, omitting the explicit time dependence, the RO takes the form
\begin{equation}
    \label{eq:R_phi}
    {
    \begin{split}
        \RO_{\psi(t)} =& \sum_\alpha\gamma_\alpha L_\alpha P_{\psi(t)} L^\dagger_\alpha\\
        &+ \frac12\big(\ket{\Phi_{\psi(t)}}\bra{\psi(t)} + \ket{\psi(t)}\bra{\Phi_{\psi(t)}}\big).
    \end{split}
    }
\end{equation}

The unraveling is obtained by considering jumps $\ket{\psi(t)}\mapsto\ket{{\varphi_{\psi(t)}^j}}$ to the eigenstates of the RO, with probability {$p^j_{\psi(t)} = \lambda_{\psi(t)}^j\,dt$}, where {$\lambda_{\psi(t)}^j$} is the corresponding eigenvalue.
The deterministic evolution is, up to normalization, given by
\begin{equation}
    \label{eq:state-dep-free-evol}
    \begin{split}
        \ket{\psi(t)}\mapsto& \ket{\tilde\psi^{\text{det}}(t+dt)} = (\1-i {K_{\psi(t)}}\,dt)\ket{\psi(t)}\\
        =&(\1-i K(t)\,dt)\ket{\psi(t)} - \frac{dt}2\,{\ket{\Phi_{\psi(t)}}.}
    \end{split}
\end{equation}

In Appendix~\ref{app:RO-ME}, we show that averaging over all the realizations indeed reproduces the master equation \eqref{eq:ME}.
The key argument is that each state $\ket{\psi_i(t)}$ evolves, on average, as it would according to \eqref{eq:ME}, for any possible choice of {$\RO_{\psi_i(t)}$}.

In Table \ref{tab:RO_comparisons}, we compare the newly introduced {$\Psi$}-ROQJ with other unraveling methods; namely, MCWF, $W$-ROQJ, $R$-ROQJ and NMQJ.
In particular, we take into account their range of applicability, as well as whether each trajectory can be realized independently from the others.

\section{Characterization and limitations}
\label{sec:characterization}
In this Section, we introduce the main features of the generalized RO, showing in particular that it is always possible to have a positive RO for any P-divisible dynamics.
More interestingly, we also discuss some necessary conditions for its positivity even when P-divisibility is broken.

\begin{figure}
    \centering
    \includegraphics[width=\linewidth]{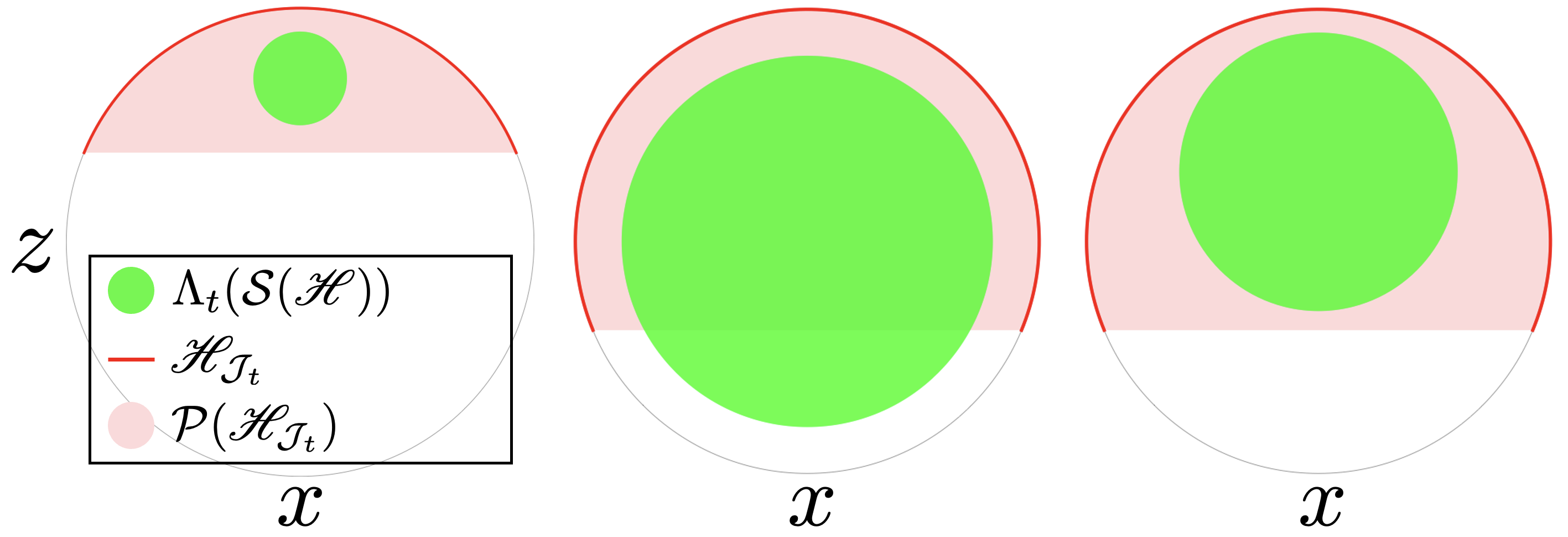}
    \caption{Examples of positivity domain \pd{} (red), convex combination of projectors on elements of the positivity domain $\mathcal P(\pd)$ (lighter red), and the time-evolved Bloch sphere $\Lambda_t(\mathcal S(\mathscr H))$ (green) for three non-P-divisible qubit dynamics at some fixed time $t$, showing the slice $y=0$ of the Bloch sphere.
    Left: no positive unraveling, since \pd{} doesn't contain any orthonormal basis.
    Middle: no positive unraveling, since $\Lambda_t\left(\mathcal S(\mathscr H)\right)\not\subseteq \mathcal P(\mathscr H_{\mathcal J_t})$.
    Right: both necessary conditions hold, therefore a positive unraveling could exist.}
    \label{fig:PD_example}
\end{figure}

\subsection{Comparison with the $W$-ROQJ unravelings}
\label{subsec:RO-W}
By using the formalism of the $W$-ROQJ of Eq.~\eqref{eq:W}, it is always possible to unravel any P-divisible dynamics by considering orthogonal jumps to the eigenstates of such operator.
It is possible to obtain the same unravelings also with the {$\Psi$}-ROQJ formalism by imposing that {$\psi(t)$} is an eigenstate of the RO: {$\RO_{\psi(t)}\ket{\psi(t)}=\lambda^{\text{det}}\ket{\psi(t)}$} for all {$\psi(t)$}, where $\lambda^{\text{det}}$ is the corresponding eigenvalue.
This way, the other post-jump states {$\ket{\varphi^j_{\psi(t)}}$} are orthogonal to {$\ket{\psi(t)}$} and it can be shown that the corresponding eigenvalues are (see Appendix~\ref{app:proof-pos-W})
\begin{equation}
    \label{eq:eigs-W}
    {
    \lambda^j_{\psi(t)} = \sum_\alpha\gamma_\alpha\lvert\braket{\varphi^j_{\psi(t)}|L_\alpha|\psi(t)}\rvert^2,}
\end{equation}
which, according to Eq.~\eqref{eq:P-div-general}, are positive for all states if and only if the dynamics is P-divisible.
This thus shows that $W$-ROQJ is a special instance of {$\Psi$}-ROQJ and thus that also the latter can be applied to any P-divisible dynamics (see Table \ref{tab:RO_comparisons}).

\subsection{Necessary conditions for a positive unraveling}
\label{subsec:cond-pos}
We now investigate when it possible to have a positive unraveling for dynamics breaking P-divisibility.
This possibility drastically simplifies the simulations since, for positive unravelings, the different realizations do not depend on each other.
For any non-P-divisible dynamics, there always exists some state $\psi$ and time $t$ such that ${\RO_{\psi}}\not\ge0$.
In fact, from the condition for P-divisibility \eqref{eq:P-div-general}, there always exists a state $\psi_\perp$, orthogonal to $\psi$, such that
\begin{equation}
    \label{eq:non_P_R}
    \braket{\psi_\perp|{\RO_{\psi}}|\psi_\perp} = \sum_\alpha \gamma_\alpha(t)\abs{\braket{\psi_\perp|L_\alpha(t)|\psi}}^2<0.
\end{equation}
Notice that this condition does not depend on the particular transformation {$\ket{\Phi_{\psi}}$} present in \eqref{eq:R_phi}, but only on $\mathcal J_t$.
This fact, however, doesn't necessarily limit the existence of a positive unraveling: it could still exist if one is able to describe the state {$\rho(t)$} only using {realizations $\ket{\psi_i(t)}$} for which ${\RO_{\psi_i(t)}}\ge0$.

Let us define the set of all states for which the RO can be positive as the positivity domain
\begin{equation}
    \label{eq:pd}
    \begin{split}
        \mathscr H_{\mathcal J_t}\coloneqq\Big\{\psi\in\mathscr H\,|&\,\forall\psi_\perp:\braket{\psi_\perp|\psi}=0, \\
        &\braket{\psi_\perp|\mathcal J_t[P_\psi]|\psi_\perp}\ge0\Big\},
    \end{split}
\end{equation}
where $\mathscr H$ is the system Hilbert space.
From Eq.~\eqref{eq:P-div-general}, it is evident that a dynamics is P-divisible if and only if $\pd=\mathscr H$ $\forall t$.
At variance with the the positivity domain introduced e.g. in \cite{Settimo2022a,Rodriguez2008a}, which refers to a subset of the set $\mathcal L(\mathscr H)$ of linear operators on $\mathscr H$, the positivity domain defined in Eq.~\eqref{eq:pd} is a subset of $\mathscr H$.
{However, $\mathscr H_{\mathcal J_t}$ is not a Hilbert space, and in particular it is not even a linear space.}

A positive unraveling can exist only if it is possible to write any state as a convex combination $\rho(t) = \sum_i p_i(t){P_{\psi_i(t)}}$ using only states $\psi_i(t)\in\mathscr H_{\mathcal J_t}$ $\forall t$.
Naturally, if $\mathscr H_{\mathcal J_t} = \emptyset$ for some time, it is not possible to have a positive unraveling.
If we further assume invertibility of the dynamical map, then it is also not possible if $\mathscr H_{\mathcal J_t}$ is zero-measured: if, for any initial state, all stochastic realizations were inside a zero-measure set, then the whole set of quantum states $\mathcal S(\mathscr H)$ would be mapped to a zero-measure set, thus breaking invertibility.

On the other hand, if $\mathscr H_{\mathcal J_t}$ is sufficiently large, we can consider using the freedom given by the generalized RO to have all realizations $\ket{\psi_i(t)}\in\mathscr H_{\mathcal J_t}$ $\forall t$.
There are two necessary conditions that $\mathscr H_{\mathcal J_t}$ must obey in order to have a positive unraveling:
\begin{enumerate}
    \item It must contain an orthonormal basis, otherwise some of states that have jumped at time $t-dt$ would violate positivity.
    \item Each state $\rho(t)$ can be written as convex combinations of elements of $\mathscr H_{\mathcal J_t}$.
    Equivalently, let $\mathcal P(\mathscr H_{\mathcal J_t})$ be the set of all convex combinations projectors $P_\psi$, with $\psi\in\mathscr H_{\mathcal J_t}$, then 
    \begin{equation}
        \Lambda_t\left(\mathcal S(\mathscr H)\right)\subseteq \mathcal P(\mathscr H_{\mathcal J_t})\quad\forall t.
    \end{equation}
\end{enumerate}
These condition are depicted pictorially in Fig.~\ref{fig:PD_example}.
Interestingly, they do not depend on the unraveling but only on the {dynamical map} $\Lambda_t$.
Therefore it is possible to rule out the possibility to have a positive unraveling for some dynamics by simply looking at $\Lambda_t$.
On the other hand, as we will see in Sec.~\ref{subsec:non-pos-unravelings} such conditions can be satisfied even by non P-divisible dynamics, which in fact do admit a positive unraveling

It is worth emphasizing that, although currently not known, the conditions for having a positive unraveling with the {$\Psi$}-ROQJ formalism cannot depend only locally on time.
In fact, the left-hand side of condition 2. does depend on the whole time evolution before P-divisibility is broken.
Therefore it could happen that for two dynamics, although having the same behavior at times when divisibility is broken, their evolution might differ at previous times and so the inclusion of condition 2. may or may not hold depending solely on the times before the violation of divisibility.
This will be illustrated explicitly in the evolution considered in Sec. \ref{sec:examples}.


\section{Control of the realizations}
\label{sec:examples}
We now proceed to present some examples of unravelings obtained using the {$\Psi$}-ROQJ formalism, showing its increased flexibility compared to the previous methods.
In particular, we demonstrate that it is possible to have positive unravelings also for dynamics violating P-divisibility.

\begin{figure*}
    \centering
    \includegraphics[width=\linewidth]{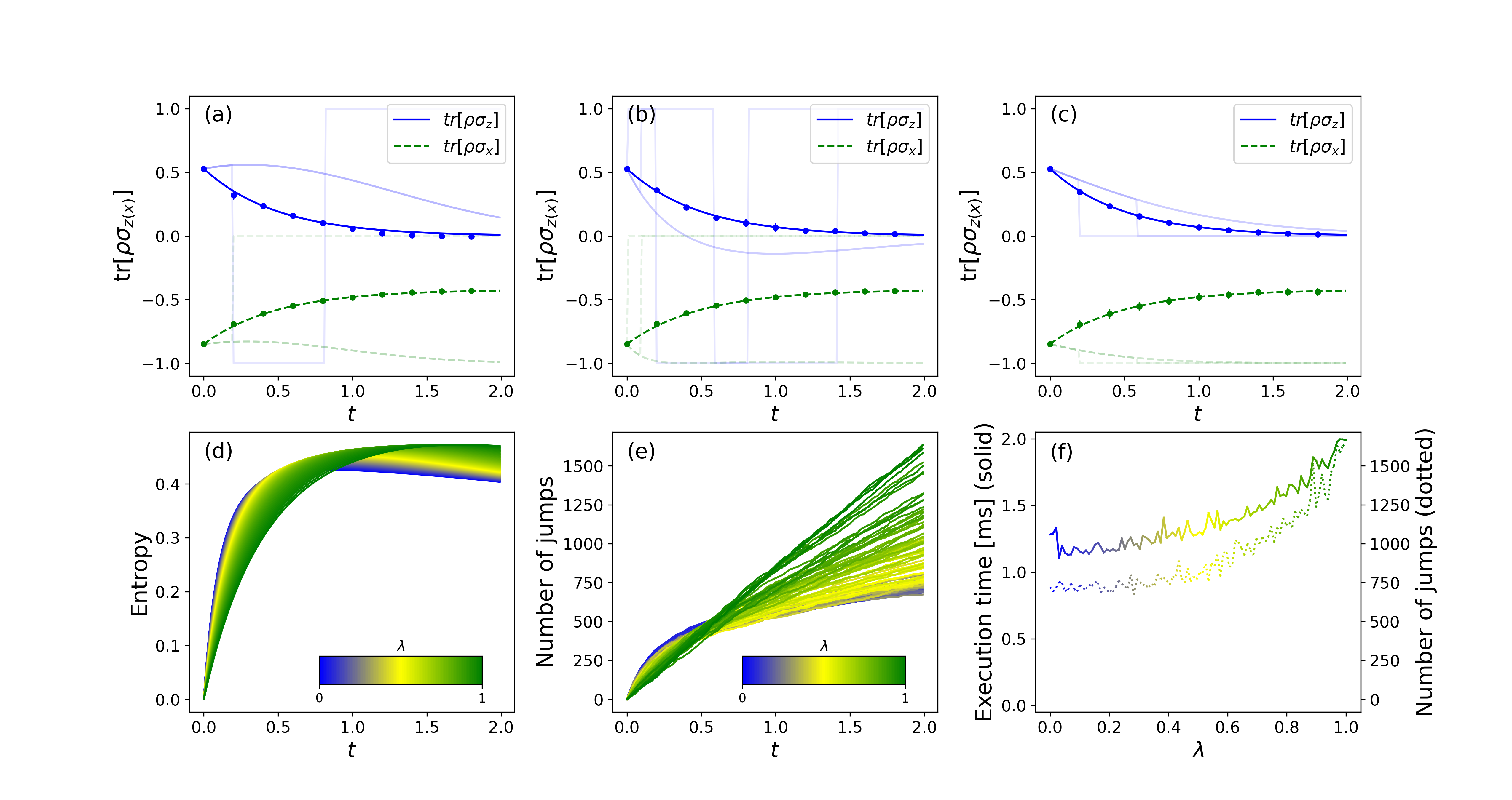}
    \caption{Eternally non-Markovian dynamics{, for the initial state $\ket{\psi(0)} = \alpha\ket0+\sqrt{1-\abs\alpha^2}\ket1$, with $\alpha\approx-0.49$.}
    Top: $z$ (blue, solid) and $x$ (green, dashed) components of the Bloch vectors. The thick solid lines are the exact results, the dots are obtained with the RO technique. In lighter shade, 5 realizations are shown. The unravelings are obtained using $N=10^3$ states. (a): $\phi_1= \phi_1^{\text{ub}}$ and only jumps to $\ket0$; (b): $\phi_1= \phi_1^{\text{lb}}$ and only jumps to $\ket1$; (c): jumps to $\ket\pm$.
    Bottom: characterization of the stochastic realizations for different choices of $\phi_1 = \lambda\phi_1^{\text{lb}}+(1-\lambda)\lambda\phi_1^{\text{ub}}$, $0\le\lambda\le1$. (d): Shannon entropy $H(\{p_i\})=-\sum_ip_i\log_2p_i$ for the probability distribution $\{p_0,p_1,p_\psi\}$ of the occupations of the states $\ket0,\ket1,\ket{\psi_{\text{det}}(t)}$; (e): number of jumps, using $N=10^4$ states; (f): computational time (left axis, solid) and total number of jumps (right axis, dotted).
    }
    \label{fig:enm-diff-unr}
\end{figure*}

\subsection{Phase covariant dynamics}
\label{subsec:ph-cov}
Let us consider a generic qubit phase covariant dynamics, i.e. a dynamics $\Lambda_t$ satisfying covariance with respect to phase transformations, namely 
\begin{equation}
    \label{eq:phase_covariance_condition}
    e^{-i\sigma_zt}\Lambda_t[\rho]e^{i\sigma_zt}=\Lambda_t[e^{-i\sigma_zt}\rho e^{i\sigma_zt}].
\end{equation}
A phase covariant dynamics has a jump term of the form \cite{Haase-fundamental, Smirne-ultimate, Vacchini2010-notes}
\begin{equation}
    \label{eq:ph_cov}
    \mathcal J[\rho] = \gamma_+\sigma_+\rho\sigma_- + \gamma_-\sigma_-\rho\sigma_++ \gamma_z\sigma_z\rho\sigma_z,
\end{equation}
where $\sigma_+=\ket1\bra0=\sigma_-^\dagger$, and a free Hamiltonian $H\propto\sigma_z$.
Such dynamics is CP-divisible if and only if all rates are positive, and P-divisible whenever \cite{Filippov-ph-cov, Teittinen2018}
\begin{equation}
    \label{eq:ph_cov_P-div}
    \gamma_\pm\ge0,\quad\text{and}\quad\gamma_z\ge-\frac12\sqrt{\gamma_+\gamma_-}.
\end{equation}

We now show that, as long as P-divisibility holds, it is always possible to have a positive unraveling using only three states: {the eigenstates $\ket0,\ket1$ of $\sigma_z$, and $\ket{\psi_{\text{det}}(t)}$, the initial state deterministically evolved up to time $t$ according to Eq.~\eqref{eq:state-dep-free-evol}.}
The possibility of using such a small effective ensemble drastically simplifies the simulations.
The RO is chosen such that $\ket{\psi_{\text{det}}(t)}$ only jumps to $\ket0,\ket1$ and, after one jump has occurred, only jumps $\ket1\leftrightarrow\ket0$ are present, without any deterministic evolution.
Thus, the effective ensemble used for the simulations only contains three states.
The possibility of considering such finite (and small) effective ensemble, significantly enhances the computational efficiency of this method, since one is not required to compute at each time-step the evolution of all states, but only needs to update the probability of occupation of such states.
This fact is particularly interesting in comparison with the NMQJ method that, expect for some special cases, needs infinitely many states in the effective ensemble.

If no jumps have occurred, it is possible to have jumps {$\ket{\psi(t)}\mapsto\ket0,\ket1$, with $\ket{\psi(t)}$ deterministically evolved according to Eq.~\eqref{eq:state-dep-free-evol},} by imposing that $\ket0$ is an eigenstate of {$\RO_{\psi(t)}$}.
This corresponds to a transformation defining the RO of Eq.~\eqref{eq:Phi} of the form
\begin{equation}
    \label{eq:Phi_ph_cov}
    \ket{\Phi_{{\psi(t)}}} = \alpha\left(2\gamma_z-\frac{\phi_1}{\sqrt{1-\abs\alpha^2}}\right)\ket{0} + \phi_1\ket1,
\end{equation}
where {$\ket{\psi(t)} = \alpha\ket0+\sqrt{1-\asq}\ket1$}, with $\alpha$ that, without loss of generality, can be assumed to be real because of phase covariance, while $\phi_1$ which can be chosen freely inside a suitable time- and state-dependent interval $\phi_1\in[\phi_1^{\text{lb}},\phi_1^{\text{ub}}]$ (for the details, the definition of $\phi_1^{\text{lb}},\phi_1^{\text{ub}}$ and the proof of the positivity, see Appendix~\ref{app:pos_ph_cov_P}).
The freedom in choosing $\phi_1$ allows us to have different realizations for the unraveling: even if the post-jump states are fixed, it is possible to unravel the dynamics with different jump rates and deterministic evolutions.
In particular, for $\phi_1 = \phi_1^{\text{lb}}$, only jumps {$\ket{\psi(t)}\mapsto\ket1$} are allowed, while for $\phi_1 = \phi_1^{\text{ub}}$, only {$\ket{\psi(t)}\mapsto\ket0$}.
On the other hand, for any $\phi_1 = \lambda\phi_1^{\text{lb}}+(1-\lambda)\phi_1^{\text{ub}}$, $0<\lambda<1$, the unraveling remains positive, with jumps to both eigenstates of $\sigma_z$ and with the possibility of choosing different jump rates depending on $\lambda$.

\subsubsection{Eternally non-Markovian dynamics}
To move further, we focus on a simple, yet significant example, namely, the eternally non-Markovian dynamics, i.e. a phase covariant dynamics with rates \cite{Hall-canonical-ME, Megier-enm}
\begin{equation}
    \label{eq:enm}
    \gamma_+(t)=\gamma_-(t)=1,\quad\gamma_z(t)=-\frac12\tanh t.
\end{equation}
The negativity of $\gamma_z$ at all times implies that such dynamics is CP-indivisible at all times, and therefore cannot be unraveled using the standard MCWF nor its non-Markovian generalization \cite{Piilo-NMQJ-PRA, Piilo-NMQJ-PRL} because of the negativity of the rate since the very beginning of time.
However, it is possible to unravel it using the generalized RO.
In addition, it is possible to realize qualitatively different realizations of the stochastic process.
In Fig.~\ref{fig:enm-diff-unr} (a)-(b), we show unravelings obtained either with only jumps to $\ket0$ or to $\ket1$ by suitably choosing $\phi_1$.
The code used for obtaining the simulations is available at \cite{github}.
Additionally, it can also be unraveled with jumps to $\ket\pm = (\ket0\pm\ket1)/\sqrt2$, by imposing $\ket\pm$ to be eigenstates of {$\RO_{\psi(t)}$}, thus giving
\begin{equation}
    \label{eq:enm_pm_jumps}
    \ket{\Phi_{{\psi(t)}}^\pm} = 2(1-\gamma_z)\sqrt{1-\abs{\alpha_-}^2}\ket+,
\end{equation}
where {$\ket{\psi(t)} = \alpha_-\ket-+\sqrt{1-\abs{\alpha_-}^2}\ket+$}, with $\alpha_-$ that can be chosen to be real because of phase covariance.
Such unraveling is shown in Fig.~\ref{fig:enm-diff-unr} (c).
This model shows the flexibility of the generalized RO, which allows us not just to have an effective ensemble consisting of only three states, but also to choose such ensemble in non-unique ways and with different deterministic evolutions for the initial state. 

The possibility of controlling the realizations in many different ways is an evident advantage of the new method proposed since it allows us to choose the most convenient ensemble, thus improving the computational efficiency of our model.
In Fig.~\ref{fig:enm-diff-unr} (d), we show the Shannon entropy $H(\{p_i\}) = -\sum_ip_i\log_2p_i$ of the probabilities $\{p_0,p_1,p_\psi\}$ of the occupations of the states $\ket0,\ket1,\ket{\psi_{\text{det}}(t)}$ for different choices of $\phi_1 = \lambda\phi_1^{\text{lb}}+(1-\lambda)\phi_1^{\text{ub}}$.
Therefore, it is possible to choose the unravelings in order to minimize the amount of classical information required to describe the average state $\rho = \sum_ip_i{P_{\psi_i}}$.
In Fig.~\ref{fig:enm-diff-unr} (e), we show that it is possible to have unravelings with vastly different number of jumps involved.
This is crucial for the computational efficiency since the fewer jumps involved the more efficient the simulation is, since one does not need to diagonalize the RO at each time-step, but only at the rare times in which a jump happens \cite{Luoma-WTD}.
This can be done by connecting the jump probability to the reduction of the norm of the deterministic state.
This fact is shown in Fig.~\ref{fig:enm-diff-unr} (f), in which we show a strong dependence between the total number of jumps, with more efficient simulations for fewer jumps.

\subsection{Positive unraveling for non-P-divisible dynamics}
\label{subsec:non-P-div}

\begin{figure*}[t!]
    \centering
    \includegraphics[width=\linewidth]{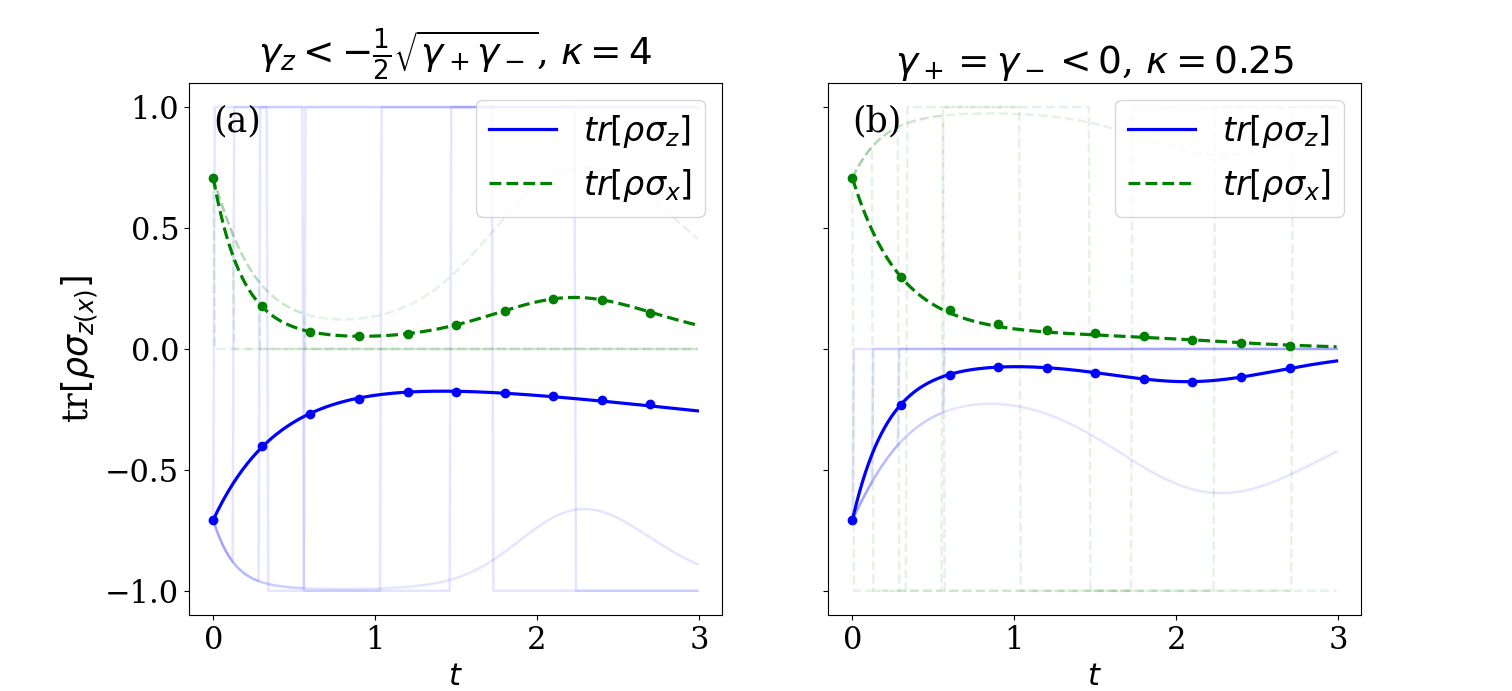}
    \caption{Positive unravelings for non-P-divisible phase covariant dynamics {for the initial state $\ket{\psi(0)} = \alpha\ket0+\sqrt{1-\abs\alpha^2}\ket1$, with $\alpha\approx0.92$.} $z$ (blue, solid) and $x$ (green, dashed) components of the Bloch vectors. The thick solid lines are the exact results, the dots are obtained with the RO technique. In lighter shade, 5 realizations are shown. (a): $\gamma_z<-\frac12\sqrt{\gamma_+\gamma_-}$ as in Eq.~\eqref{eq:no-P_ph_cov_gamma_z_rates} with $\kappa = 4$; (b): $\gamma_+=\gamma_-<0$ as in Eq.~\eqref{eq:no-P_ph_cov_gamma_pm_rates} with $\kappa = 0.25$.
    The unravelings are obtained using $N=10^4$ states.}
    \label{fig:Non-P-div_positive}
\end{figure*}

As one of our main results, we now proceed to show that it is possible to have positive unravelings for some dynamics that violate P-divisibility.
We consider again the phase covariant dynamics~\eqref{eq:ph_cov}, and we show that one can have positive unravelings when P-divisibility is broken by any of the two conditions of Eq.~\eqref{eq:ph_cov_P-div}.

\subsubsection{P-divisibility broken by $\gamma_z$}
Let us start from the case that, for some time $t$,
\begin{equation}
    \label{eq:no-P-div_gamma_z}
    \gamma_z = -\frac12\sqrt{\gamma_+\gamma_-}-\varepsilon,
\end{equation}
where $\varepsilon>0$ quantifies the violation of P-divisibility according to the second condition of Eq.~\eqref{eq:ph_cov_P-div}.
Let us consider again a RO with $\Phi_\psi$ in the form of Eq.~\eqref{eq:Phi_ph_cov}.
If, for $\gamma_z = -\frac12\sqrt{\gamma_+\gamma_-}$, any $\phi_1$ in the interval $[\phi_1^{\text{lb}},\phi_1^{\text{ub}}]$ gives a positive unraveling, the effect of $\varepsilon$ is that of shrinking the allowed interval $[\phi_1^{\text{lb}},\phi_1^{\text{ub}}]\mapsto[\phi_1^{\text{lb}}+2\varepsilon,\phi_1^{\text{ub}}-2\varepsilon]$.
But, for a sufficiently small $\varepsilon$, this new interval will still exist for all $\psi$ for which $\phi_1^{\text{lb}} \ne \phi_1^{\text{ub}}$, i.e. 
\begin{equation}
  \label{eq:fail_pos_unr_gamma_z}
  \sqrt{\frac{\gamma_-(t)}{\gamma_+(t)}} \ne \frac{1-\abs\alpha^2}{\abs\alpha^2},
\end{equation}
where $\alpha = \braket{0|\psi}$.
Therefore, a positive unraveling will still exist, provided that one is able to describe the dynamics with states $\ket{\psi_{i}(t)}$ such that this condition always holds at times for which P-divisibility is broken.
In other words, recalling the two necessary conditions of Sec.~\ref{subsec:cond-pos}, the positivity domain \pd{} is the whole Bloch sphere, excluding the state for which Eq.~\eqref{eq:fail_pos_unr_gamma_z} holds and their neighbourhood.
Therefore, since \pd{} is large, we can find unravelings for which condition 2. holds.
Also condition 1. holds, since $\ket0,\ket1\in\pd$ and therefore the positivity domain contains an orthonormal basis at all times.
We are therefore led to unravel the dynamics with jumps to such orthonormal basis.

As an example, let us consider the rates \cite{Teittinen2021}
\begin{equation}
    \label{eq:no-P_ph_cov_gamma_z_rates}
  \gamma_+(t) = e^{-t/2},\quad
  \gamma_-(t) = e^{-t/4},\quad
  \gamma_z(t) = \frac\kappa2e^{-\frac38t}\cos(2t),
\end{equation}
for which P-divisiblity is violated for $\kappa>1$.
In Figure~\ref{fig:Non-P-div_positive} (a), we show a positive unraveling for $\kappa=4$.
Again, it is possible to use an effective ensemble consisting only of 3 states $\{\ket0,\ket1,\ket{\psi_{\text{det}}(t)}\}$, thus greatly simplifying the task of simulating such non-Markovian dynamics.
It is worth noticing that this particular value of $\kappa=4$ gives positive unraveling only for some initial states, while for others the eigenvalues might become negative.
Nevertheless, there exist values $1<\kappa\le\kappa_{\text{max}}\approx1.2$ for which P-divisibility is broken and the unraveling is positive for all initial states.
In Appendix~\ref{app:proof_pos_unr_non-P}, we show the existence of such $\kappa_{\text{max}}>1$ for which a positive unraveling exists for all initial states.
For all values of $\kappa$, the resulting dynamics is qualitatively similar to the one presented in Fig.~\ref{fig:Non-P-div_positive} (a) for $\kappa=4$.

The non-Markovian behavior is evident because of the non-monotonic behavior of the coherence of $\rho(t)$.
This non-monotonicity is entirely captured by the deterministic state, which evolves towards the equator of the Bloch sphere (thus increasing its coherence) at times when P-divisibility is broken.
Therefore, a positive unraveling of this type is possible only as long as the absolute value of the recoherence for $\rho(t)$ is strictly smaller than the maximal possible recoherence for $\ket{\psi_\text{det}(t)}$ times the fraction of realizations in this state before P-divisibility is broken.
For comparison, if one would use the NMQJ, the recoherence would happen because of the reverse jumps that recreate the superposition of $\ket0$ and $\ket1$.

\subsubsection{P-divisibility broken by $\gamma_\pm$}
It is possible to have a positive unraveling also when the violation of P divisibility arises from $\gamma_\pm < 0$.
Unlike the previous case, it is not possible to have $\ket0$ or $\ket1$ in the effective ensemble because they are not in the positivity domain \pd, since
\begin{equation}
    \braket{0|\mathcal J_t[{P_1}]|0} = \gamma_-<0,
\end{equation}
and therefore $\RO$ cannot be positive for these states.
On the other hand, by imposing jumps to $\ket\pm$, it is possible to have a positive unraveling.
Let us focus, for the sake of simplicity, to the case $\gamma_+=\gamma_-\eqqcolon\gamma$.
Imposing jumps to $\ket\pm$ and solving for $\ket{\Phi_{{\psi(t)}}}$ gives us
\begin{equation}
    \label{eq:Phi_pm_non-P}
    \ket{\Phi_{{\psi(t)}}^\pm} = \frac{\sqrt{1-\abs{\alpha_-}^2}}{\alpha_-}[2\alpha_-(\gamma-\gamma_z)-\phi_-]\ket++\phi_-\ket-,
\end{equation}
with $\ket\psi = \alpha_-\ket{-}+\sqrt{1-\abs{\alpha_-}^2}$, with $\alpha_-$ that, because of phase covariance, can be taken to be real.
Furthermore, since $\gamma_+ = \gamma_-$, it is possible to have $\ket\pm$ which don't evolve deterministically but only via jumps $\ket\pm\mapsto\ket\mp$ with positive rates, thus having again a three-dimensional effective ensemble $\{\ket\pm,\ket{\psi_{\text{det}}(t)}\}$.

In Figure~\ref{fig:Non-P-div_positive} (b), we show such unraveling for the rates
\begin{equation}
    \label{eq:no-P_ph_cov_gamma_pm_rates}
    \gamma = \frac12 e^{-t/4}[\kappa + (1-\kappa)e^{-t/4}\cos(2t)],\quad
    \gamma_z=\frac12.
\end{equation}
It is possible to notice a revival in the absolute value of $\operatorname{tr}[\rho\sigma_z]$ for $t\approx 2$, which is a clear indication of the non-Markovian behavior.
It is worth noticing that, for both dynamics of Figure~\ref{fig:Non-P-div_positive}, non-Markovianity is also witnessed by a revival of the trace distance.
Therefore, the condition for a positive unraveling cannot coincide with the BLP condition  \cite{BLP,BLP-PRA} of monotonicity of the trace distance.

Figure~\ref{fig:Non-P-div_positive} (b) also shows why a generic condition for the existence of a positive unraveling cannot be local in time.
Here, before P-divisibility is broken, the $z$ component of the Bloch vector is reduced, mapping the whole Bloch ball close to the equator.
This fact is crucial since $\ket0$ and $\ket1$ (as well as a neighborhood of them) lie outside \pd, so that the only way to have a positive unraveling is to be able to describe $\rho(t)$ without using such states outside \pd.
Therefore, the second necessary condition of Sec.~\ref{subsec:cond-pos} can only hold if $\mathcal S(\mathscr H)$ is mapped towards the equator of the Bloch sphere before P-divisibility is broken.

\begin{figure*}[t!]
    \centering
    \includegraphics[width=.9\linewidth]{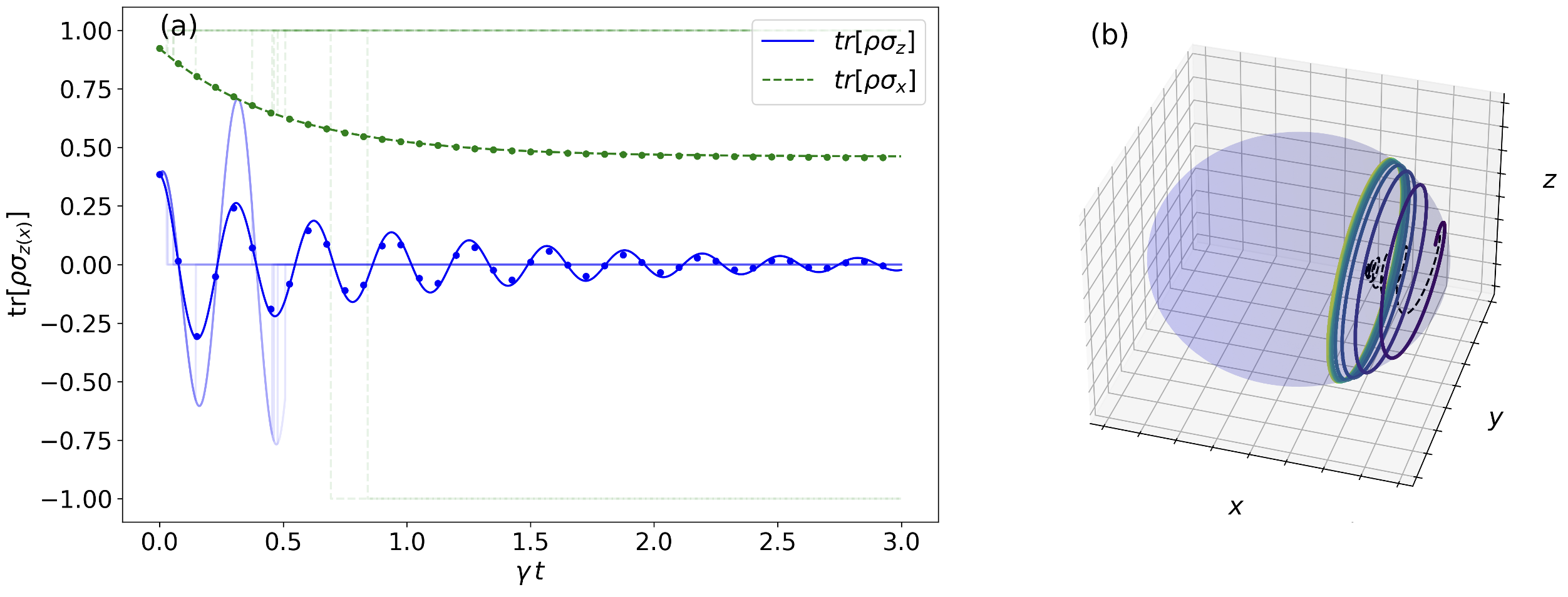}
    \caption{Driven dynamics of Eq.~\eqref{eq:ph_cov_driven_rates} for the ratio $\beta/\gamma=10$ {and initial state $\ket{\psi(0)} = \alpha\ket0+\sqrt{1-\abs\alpha^2}\ket1$, with $\alpha\approx0.20$}. (a): $z$ (blue, solid) and $x$ (green, dashed) components of the Bloch vectors. The thick solid lines are the exact results, the dots are obtained with the RO technique. In lighter shade, 5 realizations are shown.
    The unraveling is obtained using $N=10^4$ states.
    (b): trajectory in the Bloch sphere of $\rho(t)$ (black, dashed) and of $\ket{\psi_{\text{det}}(t)}$.}
    \label{fig:driven}
\end{figure*}

\subsection{Non-positive unravelings}
\label{subsec:non-pos-unravelings}
Although we have proven that there exist non-P-divisible dynamics that can be positive unraveled using the {$\Psi$}-ROQJ formalism, these positive unravelings do not exist for all non-P-divisible dynamics.
We now present some examples for which a positive unraveling using the {$\Psi$}-ROQJ does not exist, by finding dynamics for which the two necessary conditions of Sec.~\ref{subsec:cond-pos} for the existence of positive unravelings are violated.

First of all, it is easy to notice that, whenever all rates are negative at the same time, Eq.~\eqref{eq:non_P_R} is satisfied for (almost) all states, thus $\pd$ is zero-measured, which, as discussed in Sec.~\ref{subsec:cond-pos}, implies that no positive unraveling can be devised even with the generalized RO formalism.
As a special case, it is easy to see that a qubit pure dephasing dynamics $d\rho/dt = \gamma(t)(\sigma_z\rho\sigma_z-\rho)$ cannot have a positive unraveling whenever $\gamma(t)<0$, since
\begin{equation}
    \braket{\psi_\perp|{\RO_{\psi}}|\psi_\perp} = 4\abs\alpha^2(1-\abs\alpha^2)\gamma(t)<0,
\end{equation}
where $\alpha=\braket{0|\psi}$ and therefore \pd{} only contains $\ket0$ and $\ket1$.

For a generic phase covariant dynamics \eqref{eq:ph_cov}, it is possible to characterize the positivity domain by considering the infinitesimal time evolution of the state $\psi$.
Let $\mathbf r_\psi$ be the Bloch vector associated to the state, then $\psi$ is in \pd{} if and only if $\norm{\mathbf r_\psi(t)} = 1 \ge \norm{\mathbf r_\psi(t+dt)}$.
For the phase covariant, 
\begin{equation}
    \label{eq:dr_ph_cov}
    \begin{split}
        \norm{\mathbf r_\psi(t+dt)}^2 = 1-dt\big[&\gamma_++\gamma_-+4\gamma_z + 2z(\gamma_--\gamma_+)\\
        &+z^2(\gamma_++\gamma_--4\gamma_z)\big],
    \end{split}
\end{equation}
with $z = \braket{\psi|\sigma_z|\psi}$, from which it is easy to determine \pd.
For the special case $\gamma_+=\gamma_-=\gamma<0$, $\gamma_z>0$, $\psi\in\pd$ if and only if
\begin{equation}
    \label{eq:pd_ph_cov_same_rates}
    \abs z \le \sqrt{\frac{g-2}{g+2}},\qquad g = \abs\gamma/\gamma_z.
\end{equation}
For $g<2$, $\pd=\emptyset$; for $g=2$ it only contains the equator of the Bloch sphere, and for $g\to\infty$ only the poles are excluded, therefore not all phase covariant dynamics can be positively unraveled.

\subsection{Driven dynamics}

Simulating driven dynamics is a notoriously difficult task, especially when divisibility is violated.
We now show that with the {$\Psi$}-ROQJ formalism it is possible to simplify noticeably the task.
Let us consider, as an example, a phase covariant dynamics with rates
\begin{equation}
    \label{eq:ph_cov_driven_rates}
    \gamma_+(t)=\gamma_-(t)=\gamma,\quad \gamma_z(t) = -\frac\gamma2\tanh \gamma t,
\end{equation}
with $\gamma$ a positive constant, and a driving
\begin{equation}
    H=\beta \sigma_x.
\end{equation}
{The jump term, except for the constant factor $\gamma$, is the same as the eternally non-Markovian of Eq.~\eqref{eq:enm}.
However, the dynamics is not phase covariant since the driving, not being proportional to $\sigma_z$, breaks phase covariance Eq.~\eqref{eq:phase_covariance_condition}.}

It is possible to unravel such dynamics, regardless of the relative strength of the driving $\beta/\gamma$, by simply using a three-dimensional effective ensemble $\{\ket{\psi_{\text{det}}(t)},\ket+,\ket-\}$, with the driving fully captured by the deterministically evolving state $\ket{\psi_{\text{det}}(t)}$ only.

From $\ket{\psi_{\text{det}}(t)}$, it is always possible to impose jumps only to the eigenstates of the driving $\sigma_x$ (for the details see Appendix~\ref{app:RO_driven}) with positive rates as long as $\gamma\ge0$.
Interestingly, such RO does not depend on the driving, therefore this method works for any value of the ratio $\beta/\gamma$.
Furthermore, it also works for time-dependent rates and/or driving.

In Fig.~\ref{fig:driven} (a), we show the agreement between the exact solution and the unraveling with the three-dimensional effective ensemble for a strong driving $\beta = 10\gamma$.
Furthermore, in Fig.~\ref{fig:driven} (b), we show the trajectory on the Bloch sphere of the exact solution $\rho(t)$ and of the deterministically evolving state $\ket{\psi_{\text{det}}(t)}$.
As $\rho(t)$ spirals towards the asymptotic state $\rho_\infty$, $\ket{\psi_{\text{det}}(t)}$ evolves rotating on the same $x$ component of the Bloch sphere of $\rho_\infty$.
However, this is not the only possible choice for the deterministic evolution, since for different unravelings one could have qualitatively different evolutions of $\ket{\psi_{\text{det}}(t)}$.

\section{Conclusions}
\label{sec:conclusions}
In this work, we have generalized the RO approach to unravel open system dynamics by allowing the RO to explicitly depend on the current state of the realization.
We have shown that this gives us additional freedom in controlling the different stochastic realizations of the jump process, even in the case of strongly driven dynamics, which are notoriously hard to simulate, by using a small effective ensemble.
With this new method, we have also shown that one is able to choose the RO in order to optimize the simulations, i.e. by minimizing the classical entropy of the probability of occupation for the states in the effective ensemble, thus minimizing the classical information needed to describe the unravelings, or the number of jumps needed, thus minimizing the computational time.

We have also shown that it is possible to simulate some P-indivisible dynamics which are non-Markovian according to the different definitions of quantum non-Markovianity \cite{BLPV-colloquium, rivas-quantum-nm}.
This is particularly remarkable since the previous methods based on the RO do not work whenever P-divisibility is broken and since our method does not require to establish correlations among different trajectories, nor to include additional degrees of freedom.
Future work aims to explicitly characterize which dynamics can be unraveled with the generalized RO method.

In the future, we will exploit the generalized RO to study more complex dynamics, with a particular focus on higher dimensional systems, since the ability of fixing a small finite-dimensional effective ensemble would noticeably simplify the computational efficiency.
We also aim to connect these state-dependent unravelings with a proper continuous-measurement scheme.
In addition, the possibility of having a broad class of distinct unravelings could allow us to study open systems evolving under non-Hermitian Hamiltonians, when conditioned to no jumps happening.

\section*{Acknowledgements}
F.S. acknowledges support from the Magnus Ehrnrooth Foundation.
A.S. acknowledges financial support from MUR under the “PON Ricerca e Innovazione 2014-2020” project EEQU.
D.C. was supported by the Polish National Science Center project No. 2018/30/A/ST2/00837.
F.S., K.L., B.V., A.S., and J.P. thank the Toru\'n group for the hospitality received.

\bibliography{biblio}

\appendix
\section{Connection to the master equation}
\label{app:RO-ME}
We now show that the state-dependent unraveling \eqref{eq:R_phi} reproduces exactly the master equation \eqref{eq:ME}.

Suppose that, at some time $t$, the state of the system is described by $\rho(t) = \sum_ip_i{P_{\psi_i(t)}}$.
Let us focus on one particular realization $\ket{\psi_i(t)}$.
In the infinitesimal time interval $dt$, the state can evolve via a jump
\begin{equation}
    \label{eq:jump_avg}
    \ket{\psi_i(t)}\mapsto{\ket{\varphi^j_{\psi(t)}}},\quad j=1,...,d
\end{equation}
to an eigenstate of {$\RO_{\psi_i(t)}$}, with probability {$p^j_{\psi_i(t)}=\lambda^j_{\psi_i(t)}dt$}, with {$\lambda^j_{\psi_i(t)}$} being the corresponding eigenvalue.
Alternatively, it can evolve deterministically
\begin{equation}
    \label{eq:det_avg}
    \ket{\psi_i(t)}\mapsto\ket{\psi^{\text{det}}_i(t+dt)} = {\frac{(\1-i\,K_{\psi_i(t)} dt)\ket{\psi_i(t)}}{\norm{(\1-i\,K_{\psi_i(t)} dt)\ket{\psi_i(t)}}}},
\end{equation}
with probability
\begin{equation}
        {p^{\text{det}}_{\psi_i(t)} = 1-\sum_jp^j_{\psi_i(t)}=1-\tr\left[\RO_{\psi_i(t)}\right]\,dt.}
\end{equation}
It is easy to see that, at the first order in $dt$, 
\begin{equation}
    \label{eq:p_det_same}
    {\norm{(\1-i\,K_{\psi_i(t)} dt)\ket{\psi_i(t)}}^2=1-\tr\left[\RO_{\psi_i(t)}\right]\,dt = p^{\text{det}}_{\psi_i(t)}}.
\end{equation}
Therefore, the average evolution for $\ket{\psi_i(t)}$ is
\begin{gather}
    \label{eq:avg_evol_single_state}
    \begin{split}
        {P_{\psi_i(t)}}\mapsto& \,{p^{\text{det}}_{\psi_i(t)}}{P_{\psi^{\text{det}}_i(t+dt)}}\\
        &+ dt\,\sum_{j=1}^d{\lambda^j_{\psi_i(t)}{P_{\varphi^j_{\psi_i(t)}}}}\\
        =&- i\,dt{(K_{\psi_i(t)}{P_{\psi_i(t)}}-{P_{\psi_i(t)}}K_{\psi_i(t)}^\dagger)}\\
        &+dt\,{\RO_{\psi_i(t)}+{P_{\psi_i(t)}}}.
    \end{split}
\end{gather}
Therefore, ${P_{\psi_i(t)}}$ evolves, on average, as by the master equation \eqref{eq:ME}.
If one averages over all possible states $\ket{\psi_i(t)}$, the exact master equation \eqref{eq:ME} is obtained also for $\rho(t)$, thus showing that such unravelings indeed reproduce the exact dynamics.

\section{Proof of Eq.~\eqref{eq:eigs-W}}
\label{app:proof-pos-W}
In the unravelings using the $W$-operator, the jumps are to states orthogonal to the pre-jump state, with positive rates whenever the dynamics is $P$-divisible.
Analogous unravelings can also be obtained using {$\RO$} by imposing
\begin{equation}
  \label{eq:psi_eig_R}
  {\RO_{\psi(t)}}\ket{\psi(t)} = \lambda^{\text{det}}\ket{\psi(t)},
\end{equation}
which, in the basis $\{\ket{\psi_i}\}_i$, with $\ket{\psi_1}=\ket{\psi(t)}$, corresponds to the choice
\begin{equation}
  \label{eq:C_orth_jumps}
  {C_{\psi(t)}} = c_{11}P_{\psi_1} - 2\sum_{i=2}^d\braket{\psi_i|{\mathcal J}_t[P_{\psi_1}]|\psi_1}\ket{\psi_i}\bra{\psi_1},
\end{equation}
with $\operatorname{Re}c_{11}\ge - \braket{{\mathcal J}_t[P_{\psi(t)}]}_{\psi(t)}$ to ensure the positivity of $\lambda^{\text{det}}$.
Besides this constraint, $c_{11}$ is a free parameter that can be used to modify the free evolution.
Also, any additional term proportional to $\ket{\psi_i}\bra{\psi_j}$, $j\ge2$, in~\eqref{eq:C_orth_jumps} does not affect the unraveling, since ${C_{\psi(t)}}$ only acts on $\ket{\psi(t)}$.
Any eigenstate $\ket{\varphi_i}\ne\ket{\psi(t)}$ of {$\RO_{\psi(t)}$} must be orthogonal to $\ket{\psi(t)}$, therefore $\ket{\varphi_i} = (\id-P_{\psi(t)})\ket{\varphi_i}$ and
\begin{equation}
    \begin{split}
        {\RO_{\psi(t)}}\ket{\varphi_i} =& {\RO_{\psi(t)}}(\id-P_{\psi(t)})\ket{\varphi_i}\\
        =& {\lambda^i_{\psi(t)}}\ket{\varphi_i}\\
        =& {\lambda^i_{\psi(t)}}(\id-P_{\psi(t)})\ket{\varphi_i}.
    \end{split}
\end{equation}
Multiplying on the left by $(\id-P_{\psi(t)})$, we obtain that $\ket{\varphi_i}$ is an eigenstate also of $W_\psi=(\id-P_{\psi}){\mathcal J}_t[P_\psi](\id-P_\psi)$.
Therefore, $W_\psi$ and {$\RO_{\psi(t)}$} have the same eigenstates and eigenvalues
\begin{equation}
  \label{eq:eigvec_R_perp}
  {\lambda^i_{\psi(t)}} = \sum_\alpha\gamma_\alpha\lvert\braket{\varphi_i|L_\alpha(t)|\psi(t)}\rvert^2,
\end{equation}
which are positive if and only if the dynamics is P-divisible.

\section{Positive unraveling for P-divisible phase covariant dynamics}
\label{app:pos_ph_cov_P}
Starting from the phase covariant master equation, imposing $\ket0$ to be an eigenstate of the RO corresponds to $\ket{\Phi_\psi}$ as in Eq.~\eqref{eq:Phi_ph_cov}.
If no jump has occurred, it is easy to see by direct calculation that the eigenvalues $\lambda_i$, corresponding to the rates for the jumps $\ket\psi\mapsto\ket i$, are
\begin{gather}
    \lambda_1 = \alpha^2\gamma_+ + \gamma_z(1-\alpha^2)+\sqrt{1-\alpha^2}\phi_1,\\
    \lambda_0 = (1-\alpha^2)\gamma_- + 3\alpha^2\gamma_z-\frac{\alpha^2}{\sqrt{1-\alpha^2}}\phi_1.
\end{gather}
Imposing the positivity of the eigenvalues, one finds
\begin{equation}
    \begin{cases}
        \phi_1\ge -\frac{\alpha^2}{\sqrt{1-\alpha^2}}\gamma_+ - \gamma_z\sqrt{1-\alpha^2}\eqqcolon \phi_1^{\text{lb}}\\
        \phi_1\le \frac{(1-\alpha^2)^3/2}{\alpha^2}\gamma_- + 3\sqrt{1-\alpha^2}\gamma_z\eqqcolon \phi_1^{\text{ub}}.
    \end{cases}
\end{equation}
Any $\phi_1\in[\phi_1^{\text{lb}},\phi_1^{\text{ub}}]$ gives $\lambda_i\ge0$ and therefore a positive unraveling.
This is possible only provided that $\phi_1^{\text{ub}}\ge\phi_1^{\text{lb}}$.
To show that this is indeed the case, one can use the conditions for P-divisibility~\eqref{eq:ph_cov_P-div} to obtain
\begin{equation}
    \begin{cases}
        \phi_1^{\text{lb}}\le \frac12{\sqrt{\gamma_+\gamma_-(1-\alpha^2)}} - \gamma_+ \frac{\alpha^2}{\sqrt{1-\alpha^2}}\eqqcolon \tilde\phi_1^{\text{lb}}\\
        \phi_1^{\text{ub}} \ge \frac{(1-\alpha^2)^{3/2}}{\alpha^2}\gamma_- -\frac32{\sqrt{\gamma_+\gamma_-(1-\alpha^2)}} \coloneqq \tilde\phi_1^{\text{ub}}.
    \end{cases}
\end{equation}
It is easy to verify that 
\begin{equation}
    \frac{\tilde\phi_1^{\text{ub}}-\tilde\phi_1^{\text{lb}}}{\sqrt{\gamma_+\gamma_-(1-\alpha^2)}} = x+\frac1x-2,\quad x = \sqrt{\frac{\gamma_+}{\gamma_-}}\frac{\alpha^2}{1-\alpha^2},
\end{equation}
and $x+1/x\ge2$ $\forall x\ge 0$.
Therefore, it is always possible to chose $\phi_1\in[\tilde\phi_1^{\text{lb}},\tilde\phi_1^{\text{ub}}]\subseteq[\phi_1^{\text{lb}},\phi_1^{\text{ub}}]$ giving a positive unraveling.

After a jump has occurred, only the states $\ket0$ and $\ket1$ are present in the realization.
It is possible to have a positive rate for the jumps $\ket 0\leftrightarrow\ket 1$ by choosing
\begin{equation}
    \ket{\Phi_\psi^{\text{post}}} = -\gamma_z\ket\psi,\quad\psi=0,1.
\end{equation}

In order to show that the effective ensemble is indeed $\{\ket0,\ket1,\ket{\psi_{\text{det}}(t)}\}$, we also have to show that $\ket0$ does not evolve deterministically according to Eq.~\eqref{eq:state-dep-free-evol}.
This is easy to show by noticing that $\ket0$ is an eigenstate of $\Gamma$ and $\ket{\Phi_0^{\text{post}}}\propto\ket0$. 
Additionally, if there is a non-trivial Hamiltonian $H$ that does not break phase-covariance (i.e. $H\propto\sigma_z$), then $\ket0$ must be also an eigenstate of $H$.
Therefore, $\ket0$ remains fixed under the free evolution.
It is possible to prove analogously that also $\ket1$ does not evolve, thus showing that the effective ensemble is indeed three-dimensional.

\section{Proof of the positivity of the unravelings for non-P-divisible dynamics}
\label{app:proof_pos_unr_non-P}

Let us now show that it is possible to have a positive unraveling for the non-P-divisible dynamics of Eq.~\eqref{eq:no-P_ph_cov_gamma_z_rates}.
Let us start by noticing that, after a jump has occurred, the jump rates $\ket0\mapsto\ket1$ and $\ket1\mapsto\ket0$ are proportional, respectively, to $\gamma_-$ and $\gamma_+$ and therefore they are positive.

\begin{figure}
    \centering
    \includegraphics[width=\linewidth]{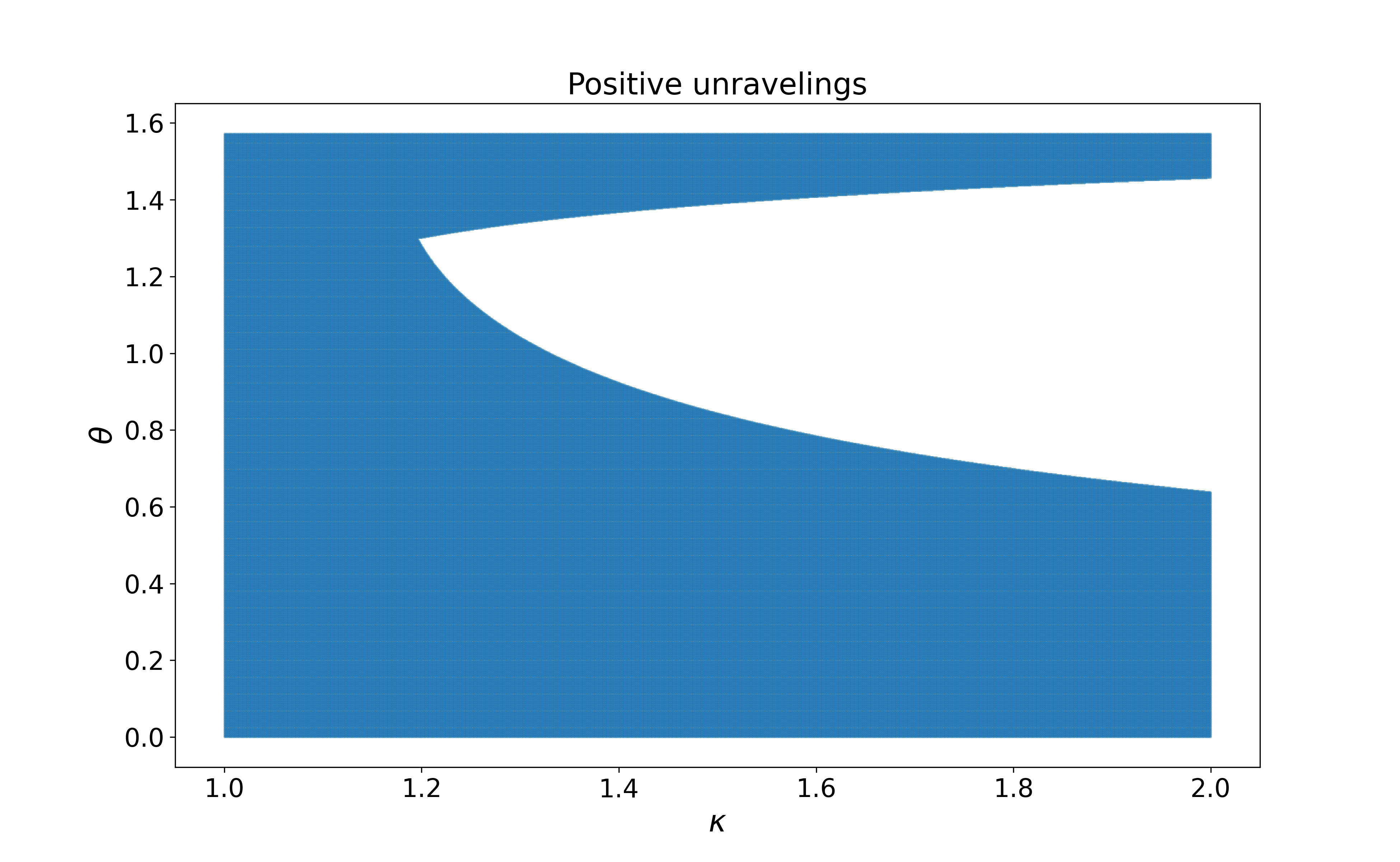}
    \caption{Positive unravelings of the dynamics of Eq.~\eqref{eq:no-P_ph_cov_gamma_z_rates} for different choices of $\kappa$ and initial states.
    The initial state is parametrized as $\ket{\psi_0} = \cos\theta\ket0 + \sin\theta\ket1$, $0\le\theta\le\pi/2$, while the transformation $\ket{\Phi_\psi}$ defining the RO as in Eq.~\eqref{eq:Phi_ph_cov}, with $\phi_1$ chosen according to Eq.~\eqref{eq:phi_1_diff_theta}.
    For all values of $\kappa\le1.2$, a positive unraveling exists for all initial states.}
    \label{fig:kappa_theta}
\end{figure}

For the deterministically evolving state, instead, there are some states for which Eq.~\eqref{eq:fail_pos_unr_gamma_z} holds, and therefore the rate would not be positive.
However, we now show that it is possible to describe the dynamics without using such states.
Let us write the initial state as $\ket{\psi_0} = \cos\theta\ket0 + e^{i\varphi}\sin\theta\ket1$.
Because of phase covariance, we can consider, without loss of generality, $\varphi = 0$ and $0\le\theta\le\pi/2$.
Let $\tilde\phi_1^{\text{lb}} \coloneqq \phi_1^{\text{lb}}+2\varepsilon$, $\tilde\phi_1^{\text{ub}} \coloneqq \phi_1^{\text{ub}}+2\varepsilon$ and fix the transformation $\ket{\Phi_\psi}$ defining the RO as in Eq.~\eqref{eq:Phi_ph_cov}.
We can use the fact that $\phi_1$ can be chosen arbitrarily inside the interval $[\tilde\phi_1^{\text{lb}}, \tilde\phi_1^{\text{ub}}]$ and choose it as
\begin{equation}
    \label{eq:phi_1_diff_theta}
    \phi_1 = \begin{cases}
        \tilde\phi_1^{\text{lb}},\quad&\text{if }\theta \ge \bar\theta\\
        \tilde\phi_1^{\text{ub}},\quad&\text{if }\theta < \bar\theta
    \end{cases}.
\end{equation}
We then sample numerically the dynamics arising from different values of $\kappa$ and $\bar\theta$ for all initial states.
In Fig.~\ref{fig:kappa_theta}, we show all positive unravelings obtained using $\bar\theta = 1.3$, for all initial states and different values of $\kappa$.
Interestingly, for all values of $\kappa\le1.2$, it is possible to have a positive unraveling for all initial states.
This value of 1.2 is only a lower bound for the true $\kappa_{\text{max}}$, but it suffices to show the existence of positive unravelings for all initial states for some non-P-divisible dynamics.
The price we have to pay to obtain this is to have a RO that depends also on the initial state, but this is by no means different from the dependence on the current state only.

\section{RO for the driven dynamics}
\label{app:RO_driven}
Let us consider, without loss of generality, $\gamma=1$.
It is possible to impose jumps $\ket{\psi_{\text{det}}(t)}\mapsto\ket{\pm}$ by choosing
\begin{equation}
    \label{eq:phi_+_driven}
    \ket{\Phi_\psi} = \frac{\sqrt{1-\asq}}{\alpha^*}\left[\alpha(1-2\gamma_z)+\alpha^*+\phi_0^*\right]\ket1+\phi_0\ket0
\end{equation}
where $\psi = \alpha\ket-+\sqrt{1-\asq}\ket+$.

The jump rates only depend on the real parameter $\xi = \phi_-\alpha^*+\phi_-^*\alpha = 2\Re\phi_-\alpha^*$ and are given by
\begin{gather}
    2\lambda_0 = (1+2\gamma_x)(1-\asq)+\asq+\xi\\
    \begin{split}
        2\asq\lambda_1 =& -\xi(1-\asq)+(1-2\gamma_x)(1-\asq)({\alpha^*}^2+\alpha^2)\\
        &+\abs\alpha^4(1+2\gamma_x)+3\asq(1-\asq)
    \end{split}
\end{gather}
and are positive whenever
\begin{equation}
    \begin{cases}
        \xi\ge-(1+2\gamma_z)(1-\asq)-\asq\eqqcolon\xi^{\text{lb}}\\
        \xi\le(1-2\gamma_z)({\alpha^*}^2+\alpha^2)+3\asq+\frac{\abs\alpha^4}{1-\asq}(1+2\gamma_z)\eqqcolon\xi^{\text{ub}}
    \end{cases}.
\end{equation}
It is easy to show that $\xi^{\text{ub}}\ge\xi^{\text{lb}}$ for any $\gamma\ge0$ or, in other words, that there always exists a solution for $\xi\in[\xi^{\text{lb}},\xi^{\text{ub}}]$.
Therefore, it is always possible to choose $\phi_-$ so that the rates for the jumps $\ket{\psi_{\text{det}}(t)}\mapsto\ket{\pm}$ are always positive.

\end{document}